# RICH METHANE PREMIXED LAMINAR FLAMES DOPED BY LIGHT UNSATURATED HYDROCARBONS PART II: 1,3-BUTADIENE

H.A. GUENICHE, P.A. GLAUDE[*], R. FOURNET, F. BATTIN-LECLERC

Département de Chimie-Physique des Réactions,

UMR 7630 CNRS, INPL-ENSIC,

1 rue Grandville, BP 20451, 54001 NANCY Cedex, France

Full-length article

SHORTENED RUNNING TITLE : **METHANE FLAMES DOPED BY 1,3-BUTADIENE**

[*] E-mail : Pierre-alexandre.glaude@ensic.inpl-nancy.fr ; Tel.: 33 3 83 17 51 01 , Fax : 33 3 83 37 81 20

In line with the study presented in the part I of this paper, the structure of a laminar rich premixed methane flame doped with 1,3-butadiene has been investigated. The flame contains 20.7% (molar) of methane, 31.4% of oxygen and 3.3% of 1,3-butadiene, corresponding to an equivalence ratio of 1.8, and a ratio $C_4H_6$ / $CH_4$ of 16 %. The flame has been stabilized on a burner at a pressure of 6.7 kPa using argon as dilutant, with a gas velocity at the burner of 36 cm/s at 333 K. The temperature ranged from 600 K close to the burner up to 2150 K. Quantified species included usual methane $C_0$-$C_2$ combustion products and 1,3-butadiene, but also propyne, allene, propene, propane, 1,2-butadiene, butynes, vinylacetylene, diacetylene, 1,3-pentadiene, 2-methyl-1,3-butadiene (isoprene), 1-pentene, 3-methyl-1-butene, benzene and toluene.

In order to model these new results, some improvements have been made to a mechanism previously developed in our laboratory for the reactions of $C_3$-$C_4$ unsaturated hydrocarbons. The main reaction pathways of consumption of 1,3-butadiene and of formation of $C_6$ aromatic species have been derived from flow rate analyses. In this case, the $C_4$ route to benzene formation plays an important role in comparison to the $C_3$ pathway.

## INTRODUCTION

With respect for the demand for much lower particulates emission from engines, especially diesel engines, an in-depth understanding of the chemistry involved in the formation of soots and polyaromatic hydrocarbons (PAH) is absolutely necessary. The chemistry leading from small unsaturated hydrocarbons to soot precursors and PAH in combustion reactive mixtures has been the subject of many studies [1], but involves some parts which are still uncertain. Different reaction pathways have been proposed for the formation and the oxidation of the first aromatic compounds, involving the reactions of $C_2$ (acetylene), $C_3$ or $C_4$ unsaturated species [2]-[6]. In a first part of this paper [7], we have investigated the reactions of allene and propyne, as they are precursors of propargyl radicals, which have an important role in benzene formation. In line with this work, it is interesting to study the reactions of 1,3-butadiene, which is a source of $C_4$ radicals and a precursor of vinylacetylene and diacetylene.

The oxidation of 1,3-butadiene has been already experimentally studied in several conditions: a shock tube [8], flow reactors [9], [10], a jet-stirred reactor [11][12], diffusion flames [13]-[14], a non-premixed co-flow flame [15]-[16], and a premixed flame [2]. The work in a laminar premixed near-sooting flame was performed with 1,3-butadiene as only fuel. The influence of the addition of 1,3-butadiene on the oxidation of methane in a flow reactor has been studied by Skjoth-Rasmussen et al. [10], and in a non-premixed co-flow flame by McEnally and Pfefferle [15] who investigated the influence of some $C_4$ compounds on the production of soot precursors. Several models have also been published to reproduce some of these experimental data [8], [13], [17]-[18].

Using the same methodology and similar experimental conditions as in part I, the purpose of the present paper is to experimentally investigate the structure of a premixed laminar methane flames containing 1,3-butadiene. That will allow comparisons with the structures of the pure methane flame and the flames doped by allene and propyne containing the same mole fractions of methane and oxygen, which has been presented in part I [7]. These results have been used to improve the mechanism previously developed in our laboratory for the reactions of 1,3-butadiene and related species [8], [19].

**EXPERIMENTAL RESULTS**

In line with our previous study [7], a laminar premixed flat flame has been stabilized on the burner at 6.7 kPa with a gas flow rate of 3.32 l/min corresponding to a gas velocity at the burner of 36 cm/s at 333 K and mixtures containing 42.8% of argon, 20.7% of methane, 33.1% of oxygen and 3.3% of 1,3-butadiene corresponding to an equivalence ratio of 1.8. The same apparatus, the same method to measure temperature, and the same analytical techniques as what is extensively described in the part I of this paper [7] have been used and they are not presented here again.

Detected species containing 3 or less carbon atoms were the same, but with different concentrations, as in the case of flames doped by allene or propyne [7] and have been separated on a Carbosphere packed column by FID and TCD; differences were encountered for heavier hydrocarbons which were analysed on a Haysep packed column by FID and nitrogen as gas carrier gas. The identification of these compounds was performed using GC/MS and by comparison of retention times when injecting the product alone in gas phase. Figure 1 presents a typical chromatogram of $C_1$-$C_6$ compounds obtained for the flame doped with 1,3-butadiene. The baseline increases because of the oven temperature program without affecting the measurements. The observed $C_4$ compounds were 1,3-butadiene (1,3-$C_4H_6$), which was the reactant and the peak of which was so large that it masked those of iso-butene or 1-butene which had been previously observed, 1,2-butadiene (1,2-$C_4H_6$), vinylacetylene ($C_4H_4$), diacetylene ($C_4H_2$) and butynes (1-$C_4H_6$ and 2-$C_4H_6$). The separation between the peaks of diacetylene, vinylacetylene and butynes is acceptable, but the peaks of 1-butyne and 2-butyne cannot be distinguished. Four $C_5$ species were also identified by GC-MS: 2-methyl-1,3-butadiene (isoprene, $iC_5H_8$), 3-methyl-1-butene ($iC_5H_{10}$), 1,3-pentadiene (1,3-$C_5H_8$) and 1-pentene (1-$C_5H_{10}$). The peaks of the two linear $C_5$ species cannot be distinguished. Two aromatic compounds are quantified, benzene ($C_6H_6$) and toluene ($C_7H_8$). Toluene cannot be seen on Figure 1 because of its longer retention time. Diacetylene, $C_5$-species and toluene were not detected in the flames doped by allene or propyne [7].

FIGURE 1

Water and small oxygenated compounds, such as acetaldehyde, formaldehyde, acroleine, were detected by GC-MS but not quantitatively analysed because of difficulties in the calibration.

1,3-butadiene (99.5% pure) and methane (99.95 % pure) were supplied by Alphagaz - L'Air Liquide. Oxygen (99.5% pure) and argon (99.995% pure) were supplied by Messer. Chromatographic analysis showed that 1,3-butadiene contained also allene (0.05%), propyne (0.05%), propene (0.2%) and 1,2-butadiene (0.05%).

Figure 2 displays the experimental temperature profiles obtained with and without the probe showing as previously that the presence of the probe induced a thermal perturbation involving a lower temperature. Without the probe, the lowest temperatures measured the closest to the burner were around 600 K. The highest temperatures were reached between 0.7 and 0.9 cm above the burner and were around 2150 K, i.e. around 200 K higher than in our previous flames (the maximum temperature was around 1850 K in the pure methane flame and around 1880 K in the flames doped by $C_3$ compounds). The temperature decreased thereafter because of the heat losses. Even if the adiabatic flame temperature is slightly lower in the case of the $C_4$ doped flame than in the $C_3$ doped flames, the maximum temperature is higher: the flame front stabilizes further from the burner and then the heat losses are decreased.

FIGURE 2

Figures 3 and 4 present the profiles of the $C_0$-$C_2$ species involved in the combustion of methane vs. the height above the burner. In the flame doped with 1,3-butadiene, the consumption of methane (fig. 3a) and oxygen (fig. 3b) and the position of the maximum concentrations of for carbon monoxide (fig. 3c) and $C_2$ compounds (fig. 4b, 4c, 4d) occur further from the burner than in the flames containing propyne and allene [7] and still more further compared to the pure methane flame. The profile of carbon dioxide (fig. 3d) shows a marked inflexion point as in the propyne and allene flame.

FIGURES 3 AND 4

As in the reference pure methane flame and in the flames doped with allene and propyne [7], ethane (fig. 4d) is experimentally produced promptly and reaches its maximum concentration close to the burner, around 0.4 cm. The profile of ethylene (fig. 4c) peaks around 0.5 cm, that of acetylene (fig. 4b) around 0.6 cm and that of carbon monoxide around 0.7 cm. While the maximum value reached by the mole fraction of ethane is not much affected by the addition of 1,3-butadiene, those of ethylene (0.007 compared to 0.002 in the pure methane flame) and acetylene (0.01 compared to 0.001) are strongly increased.

Figure 5 presents the profiles of the observed $C_3$ products, which all peak around 0.4 cm above the burner. While the formation of propane (fig. 5d) is close to that observed in the pure methane flame, the formation of allene (fig. 5a), propyne (fig. 5b) and propene (fig. 5c) are much increased by the presence of the additive. The formation of propene is higher than in the $C_3$ doped flames. This figure presents also the profiles of benzene (fig. 5e) and toluene (fig. 5f), which peak at the same location, around 0.5 cm above the burner. The maximum value reached by the mole fraction of benzene is around twice that measured in the flame doped with allene.

FIGURE 5

Figures 6 and 7 present the profiles of $C_4$ and $C_5$ species, respectively. As methane, 1,3-butadiene is consumed in the first stage of the flame, but the total consumption of the $C_4$ reactant occurs closer to the burner, at 0.5 cm height, while some methane remains up to 0.7 cm. Vinylacetylene (fig. 6d) is produced promptly and reaches its maximum concentration close to the burner, around 0.2 cm. The profiles of 1,2-butadiene (fig. 6b) and of 3-methyl-1-pentene (fig. 7b) peaks around 0.3 cm, those of butynes (fig. 6c) and isoprene (fig. 7a) around 0.35 cm, those of diacetylene (fig. 5e) and linear $C_5$ (fig. 7c) species around 0.4 cm. The maximum value reached by the mole fractions of branched $C_5$ compounds are much higher than that of the linear ones.

FIGURES 6 AND 7

## DESCRIPTION OF THE PROPOSED MECHANISM

This mechanism is an improvement of our previous mechanism that was built to model the oxidation of $C_3$-$C_4$ unsaturated hydrocarbons [7], [8], [19] to better take into account the reactions of allene, propyne and propargyl radicals. The whole mechanism involves 154 species reacting in 1055 reactions and is available on request.

*Reaction base for the oxidation of $C_3$-$C_4$ unsaturated hydrocarbons [7], [8], [19]*

This $C_3$-$C_4$ reaction base was built from a review of the recent literature and is an extension of our previous $C_0$-$C_2$ reaction base [20]. This $C_0$-$C_2$ reaction base includes all the unimolecular or bimolecular reactions involving radicals or molecules including carbon, hydrogen and oxygen atoms and containing less than three carbon atoms. The kinetic data used in this base were taken from the literature and are mainly those proposed by Baulch *et al.* [21] and Tsang *et al.* [22]. The $C_0$-$C_2$ reaction base was first presented by Barbé *et al.* [20] and has been up-dated [8].

The $C_3$-$C_4$ reaction base includes reactions involving $C_3H_2$ (CH≡CCH••), $C_3H_3$ (CH≡CCH$_2$•),

$C_3H_4$ (allene and propyne), $C_3H_5$ (3 isomers ($aC_3H_5$: •$CH_2CH=CH_2$, $sC_3H_5$: $CH_3CH=CH$•, $tC_3H_5$: $CH_3C$•$=CH_2$)), $C_3H_6$, $C_4H_2$, $C_4H_3$ (2 isomers ($nC_4H_3$: •$CH=CHC\equiv CH$, $iC_4H_3$: $CH_2=C$•$-C\equiv CH$)), $C_4H_4$, $C_4H_5$ (5 isomers ($nC_4H_5$: •$CH=CHCH=CH_2$, $iC_4H_5$: $CH_2=CHC$•$=CH_2$, $C_4H_5$-1s: $CH_3CH$•$-C\equiv CH$, $C_4H_5$-1p: $CH_2$•$CH_2C\equiv CH$, $C_4H_5$-2: $CH_2$•$C\equiv CCH_3$)), $C_4H_6$ (1,3-butadiene, 1,2-butadiene, methyl-cyclopropene, 1-butyne and 2-butyne), as well as the formation of benzene. Pressure-dependent rate constants follow the formalism proposed by Troe [23] and efficiency coefficients have been included. This reaction base was built in order to model experimental results obtained in a jet-stirred reactor for methane and ethane under atmospheric conditions [20], profiles in low pressure laminar flames of methane, acetylene and 1,3-butadiene [8] and shock tube auto-ignition delays for acetylene, propyne, allene, 1,3-butadiene [8], 1-butyne and 2-butyne [19]. An improved version has recently been used to model structure of laminar premixed flames of methane doped with allene and propyne [7].

Thermochemical data are estimated by the software THERGAS developed in our laboratory [24], which is based on the additivity methods proposed by Benson [25].

*Reactions related to 1,3-butadiene, $C_4$ and $C_5$ species*

The sub-mechanism, described below and displayed in Table I, is included in a mechanism which also contains the two reactions bases described above and which can be used to run simulations using CHEMKIN II [26]. In order to correctly model the consumption of benzene and toluene, our recent primary and secondary mechanisms for the oxidation of these species [27], [28] have also been added.

**TABLE I**

The reaction of species lighter than butadienes have been slightly up-dated. The reactions of diacetylene have not been modified except from the addition of the bimolecular initiation with oxygen molecules (reaction 6 in table 1) proposed by Hidaka et al. [32]. The rate constants of the disproportionations of $nC_4H_3$ and $iC_4H_3$ radicals with H atoms and OH radicals (reactions 11, 14, 24, 26) have been estimated from that of vinyl radicals taking into account the number of abstractable H-atoms as proposed by Wang et al. [6]. Values proposed by Wang et al. [6] have also been considered for the rate constants of the other channels of the reactions of $nC_4H_3$ and $iC_4H_3$ radicals with H atoms (reactions 10, 22). Reactions of $iC_4H_3$ and vinyl radicals have been added (reactions 28-29). The rate constants of the reaction of vinylacetylene with H atoms (reactions 34-35) have been up-dated and the reaction of vinylacetylene with acetylene to give phenyl radicals and H-atoms (reaction 42) has been added as proposed by Benson [35]. The reactions of $nC_4H_5$ and $iC_4H_5$ radicals with oxygen molecules include now the formation of both aldehydes and oxygenated radicals (reactions 62, 70) and vinylacetylene and $HO_2$ radicals (reactions 63-71), with rate constants estimated from that of vinyl radicals. This new set of rate constant favors the formation of aldehydes and should decrease the formation of vinylacetylene, which was strongly overestimated in the modeling of the flames doped by $C_3$ compounds. The combinations between $nC_4H_5$ (reaction 51) and $iC_4H_5$ (reaction 68) radicals with methyl radicals have been added.

The reactions of 1,3-butadiene have been kept unchanged, only additions of H-atoms and $CH_3$ radicals to the double bonds are now considered (reactions 78, 79, 82, 83). The rate constant of the addition of methyl radicals to 1,3-butadiene to give a linear $C_5$ species (reaction 82) is that proposed by Perrin et al. [38]. For the other additions, for which no direct data is given in the literature, the rate constants have been estimated as twice that of the similar reactions of propene [40]. The unimolecular decomposition of 1,2-butadiene to give $C_3H_3$ and $CH_3$ radicals (reaction 101) has been revisited using the software KINGAS [39], because the activation energy proposed by Leung and Lindstedt (59.5 kcal/mol [29]) was too weak compared to the enthalpy of reaction. For the addition of H-atoms to 1,2-butadiene, we have considered the formation of the three possible butenyl radicals (reactions 104-106).

It is worth noting that, as proposed by Westmoreland et al. [3], all the reactions between $C_2$ species and n-$C_4H_3$ (reactions 15-18), n-$C_4H_5$ radicals (reactions 53-60) or 1,3-butadiene molecules (reactions 86-91) and leading to aromatic and linear $C_6$ species have been considered. New reactions of cyclic $C_6$ compounds have been considered: the dehydrogenation of 1,4-cyclohexadiene to give benzene with a rate constant proposed by Ellis and Freys [44] (reaction 156), the H-abstractions from 1,4-cyclohexadiene by H-atoms and OH radicals to give $C_6H_7$ radicals with rate constants proposed by Dayma et al. [45] reactions 157 and 158) and the additions of H-atoms to benzyne ($C_6H_4$) to give phenyl radicals with a rate constant proposed by Wang et al. [6] (reaction 159).

The additions of hydrogen atoms and methyl radicals to 1,3-butadiene, lead to the formations of butenyl (reactions 78-79) and pentenyl (reactions 82-83) radicals, respectively. The reactions of these $C_4$ and $C_5$ radicals had to be added. We have considered 5 different isomers of linear butenyl radicals, the formulae of which are given in Table 1; we have considered all the possible isomerizations between these radicals (reactions 113, 114, 123, 125, 126, 128), their decomposition by beta-scission (reactions 115, 124, 127, 129) and the termination steps with H-atoms (combination and disproportionation), $CH_3$ and $HO_2$ radicals of the resonance stabilized ones (reactions 118-122). As 2-butene and 2-pentene were not detected in our experimental analysis, we have only considered reactions of the mesomer form including a secondary radical, the probability of formation of which is certainly higher than that of the mesomer form involving a primary radical.

As shown in Table 1, a linear resonance stabilized and a branched pentenyl radicals are obtained by addition of methyl radicals to 1,3-butadiene. We have considered the decomposition of these radicals by beta-scission (reactions 139, 143-145), the isomerisation of the obtained branched pentenyl radicals to give the resonance stabilized branched pentenyl radicals $iC_5H_9Y$ (reaction 142) and the combinations (reactions 140, 146) and the disproprtionnations (reactions 141, 147) of both resonance stabilized radicals with H-atoms. The disproportionations of the very stable branched resonance stabilized $iC_5H_9Y$ radicals with allyl radicals (reaction 148) and with themselves (reaction 149) have also been written. We have only considered the formation of the $C_5$ molecules observed in our experiments: 1-pentene $C_5H_{10}$, 3-methyl-1-butene $iC_5H_{10}$, 1,3-pentadiene $C_5H_8$ and isoprene $iC_5H_8$. Metathesis involving the abstraction of an allylic H-atoms have been written for these molecules (reactions 133-138, 150-155). The obtained linear and branched resonance stabilized pentadienyl radicals can react by decomposition by beta-scission (reactions 131) or by combinations with H-atoms (reactions 130-132). The formation of trienes or $C_6$ compounds has not been considered.

Apart from reaction 126, activation energy (E) for isomerisations is set equal to the sum of the activation energy for H-abstraction from the substrate by analogous radicals ($E_{abs.}$) and the strain energy of the cyclic transition state ($E_{cycle}$) and the A-factor (A) is mainly based on the changes in the number of internal rotations as the reactant moves to the transition state; the equation uses a mean value of 3.5 cal mol$^{-1}$ K$^{-1}$ for each lost rotor [37]:

$$A = e \times \frac{k_B T}{h} \times \text{rpd} \times \exp\left[\frac{(\Delta n^{\neq}_{i.rot.}) \times 3.5}{R}\right] \quad s^{-1}$$

with : $\Delta n_{int. rot.}$ : Change in the number of internal rotations as reactant moves to the transition state,

e : Base of natural logarithms,

h : Planck constant (6.6260755 10$^{-34}$ J.s$^{-1}$),

$k_B$ : Boltzmann constant (1.380658 10$^{-23}$ J.K$^{-1}$),

R : Gas constant (cal mol$^{-1}$ K$^{-1}$),

rpd :       Reaction path degeneracy = number of identical abstractable H atoms,

T :         Temperature (K).

Table 2 presents the calculation of these rate coefficient for the isomerisations of butenyl and pentenyl radicals.

**TABLE II**

The other rate constants used are mainly derived from the values given by Tsang for propene [40] or from the proposed correlations by Heyberger et al. for alkenes [41], [42].

**COMPARISON BETWEEN EXPERIMENTAL AND SIMULATED RESULTS**

Simulations were performed using PREMIX from CHEMKIN II [24] taken into account the presence of hydrocarbon impurities in 1,3-butadiene. To compensate the perturbations induced by the quartz probe and the thermocouple, the temperature profile used in calculations is an average between the experimental profiles measured with and without the quartz probe, shifted 0.15 cm away from the burner surface, as shown in figure 2.

Figures 3 and 4 show that the model reproduces satisfactorily the consumption of reactants and the formation of the main $C_0$-$C_2$ products related to the consumption of methane in the flame doped with 1,3-butadiene.

To decouple the effect due to the increase of equivalence ratio ($\Phi$) and that induced by the presence of 1,3-butadiene, figures 3 and 4 display also the results of a simulation performed for a flame containing 20.7% methane and 23.0% oxygen (with no $C_4$ additive) for $\Phi$= 1.8, i.e. equal to that of the doped flame. As the temperature rise is mainly influenced by $\Phi$, we have used the same temperature profile as to model the doped flame. As in the previous study, the profiles of methane are very similar for the doped and undoped flames at the same $\Phi$. The profiles of oxygen, carbon oxides and hydrogen are different in the doped and pure methane flames at Φ=1.8 due to a difference in the C/O and C/H ratios. At the same equivalence ratio, the maximum of ethane mole fraction is lower in the doped flame than in the pure methane flame; that could indicates a channel involving 1,3-butadiene and consuming methyl radicals, the recombination of which is the major channel of formation of ethane. The maximum of ethylene mole fraction is slighted increased by the addition of 1,3-butadiene, while that of acetylene is almost multiplied by a factor 2. As in the case of the $C_3$ additives, there is then a specific way of formation of acetylene involving 1,3-butadiene.

The simulated profiles of $C_3$ compounds and aromatic compounds (benzene, toluene) are displayed in Figure 5. The profiles of allene, propene and propane are well reproduced by simulations, while the maximum of the mole fraction of propyne is underestimated by a factor around 3. Comparison with the simulation of a pure methane flame at Φ=1.8 shows that allene, propyne and propene are much more abundant (factors around 10) in the doped flame, while the maximum of propane mole fraction is lower. Specific reactions leading to unsaturated $C_3$ products are then also induced by the presence of the $C_4$ additive. The decrease of the concentration of propane in the doped flame could confirm the possibility of a channel involving 1,3-butadiene and consuming methyl radicals (the main way of formation of propane is the combination between ethyl and methyl radicals). The profiles of benzene and toluene are also well simulated. The mole fraction of $C_4$, $C_5$ and aromatic species predicted in the pure methane flame at Φ=1.8 are always lower than $1 \times 10^{-6}$.

Figures 6 and 7 present the comparison between experimental and simulated data for $C_4$ and $C_5$ species respectively. The consumption of 1,3-butadiene is satisfactorily reproduced, as well as the profiles of 1,2-butadiene, vinylacetylene and $C_5$ species. The maximum mole fraction of butynes is underpredicted by a factor 2 and that of diacetylene overpredicted by the same factor. Simulations show that similar amounts of 1-butyne and 2-butyne are obtained and that the mole fraction of 1-pentene is about 10 times higher that that of 1,3-pentadiene.

**DISCUSSION**

Figure 8 displays the flows of consumption of the 1,3-butadiene at a temperature about 980 K corresponding to a 47% conversion. Under these conditions, 1,3-butadiene is mainly consumed by additions of H-atoms (67 %) to give butenyl radicals (1-$C_4H_7$Y and 1-$C_4H_7$-1 in Table I), of O-atoms (3%) to produce allyl radicals, H-atoms and carbon monoxide and of OH radicals (18%) to form acetaldehyde and vinyl radicals or formaldehyde and allyl radicals. While a small fraction of 1-$C_4H_7$-1 radicals leads to 1-pentene by combination with methyl radicals or to 1-$C_4H_7$Y radicals by isomerisation, the main reaction of these radicals is a beta-scission decomposition to give vinyl radicals and ethylene. This reaction explains the increase of the mole fraction of this last compound observed in figures 4c when 1,3-butadiene is added. Vinyl radicals, which are obtained both from 1,3-butadiene by OH addition and by decomposition of 1-$C_4H_7$-1 radicals, react mainly with oxygen molecules and yield acetylene that increases the amount of acetylene in the doped flame. Diacetylene is obtained from the addition of $C_2H$ radicals to acetylene. Resonance stabilized 1-$C_4H_7$Y radicals react mainly by termination steps and are the major source of 3-methyl-1-butene by combination with methyl radicals, of 1,2-butadiene by disproportionation with H-atoms and 1-butene (which cannot be analysed in our study) by combination with H-atoms. The combination of 1-$C_4H_7$Y and methyl radicals accounts for 26% of the consumption of this species and explains then the decrease in the formation of ethane and propane observed in the doped flames. The combination and the disproportionation of the resonance stabilized allyl radicals with H-atoms are the main ways of formation of propene and allene, respectively. A small fraction of allene and propyne are formed from 1-$C_4H_7$Y and 2-$C_4H_7$ radicals, respectively, which are both obtained by H-abstraction from 1,2-butadiene. The additions of methyl radicals have a very small flow rate (0.2 % of the consumption of 1,3-butadiene) and are then a minor source of $C_5$ compounds.

FIGURE 8

1,3-butadiene is also consumed by H-abstraction to give i-$C_4H_5$ radicals (6 % of its consumption) and n-$C_4H_5$ radicals (1.5). These radicals react mainly with oxygen molecules to give aldehydes and oxygenated radicals and to a much lower amount of vinylacetylene. The new rate constants that we have considered here for these reactions allow us to better reproduce the profile of vinylacetylene in the flames doped with propyne and allene, as shown in Figure 9a and increase slightly the calculated amount of benzene in the $C_3$ flames (Figure 9b) Addition of H-atoms to vinylacetylene leads to other isomers of $C_4H_5$ radicals, which can give butynes. Minor channels from i-$C_4H_5$ and n-$C_4H_5$ radicals include the combination with methyl radicals to form isoprene and 1,3-pentadiene, respectively.

FIGURE 9

A last minor channel of consumption of 1,3-butadiene is the addition/cyclization of vinyl radicals to produce cyclohexadiene (0.14 % of the consumption of 1,3-butadiene). Figure 10 presents a flow rate analysis for the production and consumption of aromatic compounds. Benzene is rapidly formed from this cyclic diene either by dehydrogenation or by metatheses with H-atoms and OH radicals, followed by the decomposition of the obtained cyclic $C_6H_7$ radicals. The $C_2$+$C_4$ route is the main benzene production channel in our conditions while the $C_3$+$C_3$ route was also of importance in some studies on pure butadiene [12][17]. This route involving vinyl radical is not commonly proposed in the formation of aromatic rings but is noted as a minor pathway yielding benzene in a $C_2H_2$ premixed flame studied by Westmoreland et al.

[3] who proposed the kinetic data used in our mechanism and by Linstedt and Skevis [17] in a pure butadiene flame. These data were based on high pressure rate constants from the literature; thermochemical assumptions and quantum-RRK calculations had been then performed to calculate rate constants at 1 atm and 50 Torr. More recently, Cavallotti et al. [46], in an ab initio study of the reactions of butadiene leading to soot formation, proposed a rate constant for the addition of $C_2H_3$ to butadiene and the subsequent decompositions or cyclization reaction of the adduct. These values lead to a higher rate of formation of cyclohexene than that of Westmoreland et al. [3] and confirm that this channel can play a role in these conditions. The high amount of vinyl radicals in the doped methane flames can explain also the importance of this channel in comparison with pure flames of unsaturated compounds. The simulated maximum amount of cyclohexadiene produced through this route is around 5 ppm, but no experimental quantification nor detection has been possible since many very small peaks of $C_5$ compounds were detected by chromatography and not well separated.

Benzene is mainly consumed by addition of O-atoms to form phenoxy radicals and by H-abstraction with H-atoms and OH radicals to give phenyl radicals. Resonance stabilized phenoxy radicals react by combination with H atoms to give phenol or by decomposition to give carbon monoxide and resonance stabilized cyclopentadienyl radicals, which lead to cyclopentadiene. About 20% of the formation of phenyl radicals is also due to the combination of propargyl radicals. Phenyl radicals are consumed by combination with methyl radicals to give toluene (65 % of its consumption) and by reactions with oxygen molecules (30 % of its consumption) to give O atoms and phenoxy radicals or H atoms and benzoquinone, which decomposes to form cyclopentadienone and carbon monoxide. Toluene react either by ipso-addition of H-atoms to give benzene and methyl radicals or by H-abstraction with H-atoms and OH radicals to form resonance stabilized benzyl radicals, which mainly decompose to produce acetylene and resonance stabilized cyclopentadienyl radicals or react with O-atoms to form benzaldehyde. H-abstractions from benzaldehyde gives carbon monoxide and phenyl radicals. Figure 9b shows that this new mechanism, including the formation of toluene, leads to higher maxima for the mole fraction of benzene in the flames doped with allene and propyne than the previous mechanism, but still keeps an acceptable agreement with experimental results In this case, an important fraction of phenyl radicals obtained by recombination of propargyl radicals is consumed by recombination with methyl radicals to give toluene molecules, which by ipso-addition of H-atoms produce benzene and methyl radicals.

FIGURE 10

**CONCLUSION**

This paper presents new experimental results for rich premixed laminar flames of methane seeded with 1,3-butadiene, as well as some improvements made to the mechanism previously developed in our laboratory for the reactions of $C_4$ unsaturated hydrocarbons. Profiles of temperature have been measured and profiles of stable species have been obtained for 24 products, including benzene, toluene and $C_3$, $C_4$ and $C_5$ unsaturated compounds. The use of methane as the background and consequently of a flame rich in methyl radicals favors the formation of $C_5$ compounds from the $C_4$ compounds.

The presence of 1,3-butadiene promotes the formation of ethylene and acetylene. The increase of the formation of this $C_2$ compounds is due to the decomposition of a butenyl radical obtained by addition of H-atoms to 1.3-butadiene. The increase in the formation of acetylene can be also attributed to the addition of OH radicals to 1,3-butadiene.

The presence of 1,3-butadiene is also responsible for the formation of benzene and toluene,

which cannot be detected in the pure methane flame. A particular interest of this flame is that the production of benzene is mainly due to reactions of $C_4$ compounds (reaction between 1,3-butadiene and vinyl radicals) and not only to the $C_3$ pathway as it is usually the case in the flames of the literature [44]. This study has also allowed us to underline the role of the combination of methyl and phenyl radicals giving toluene in the formation of benzene, through the ipso-addition of H-atoms.

To finish completely this work, a study of flames doped with cyclopentene is in progress using the same methodology.

**TABLE 1: REACTIONS OF 1,3-BUTADIENE AND OF DERIVED LINEAR UNSATURATED $C_4$ AND $C_5$ SPECIES**

The rate constants are given ($k = A\, T^n \exp(-E_a/RT)$) in cc, mol, s, kcal units. Reference numbers are given in brackets when they appear for the first time. The reactions in bold have been added or involve a modified rate constant compared to our last mechanism [7,8]. The letter V in a species name means a vinylic free radical, Y an allylic free radical and # a ring structure.

| Reactions | A | n | $E_a$ | References | No |
|---|---|---|---|---|---|
| **Reactions of C4H2 (CH≡CC≡CH, diacetylene)** | | | | | |
| C2H+C2H = C4H2 | $1.8 \times 10^{13}$ | 0.0 | 0.0 | Tsang 86[22] | (1) |
| 2C2H2 = C4H2+H2 | $1.5 \times 10^{13}$ | 0.0 | 42.7 | Leung 95[29] | (2) |
| C2H2+C2H => C4H2+H | $9.0 \times 10^{13}$ | 0.0 | 0.0 | Baulch 94[21] | (3) |
| C4H2+O = C3H2+CO | $2.7 \times 10^{13}$ | 0.0 | 1.72 | Warnatz 84[30] | (4) |
| C4H2+OH = CHO+C3H2 | $6.7 \times 10^{12}$ | 0.0 | -0.4 | Perry 84 [31] | (5) |
| **C4H2+O2 = HCCO+HCCO** | **$9.6 \times 10^{12}$** | **0.0** | **31.1** | **Hidaka02[32]** | **(6)** |
| C4H2+C2H => C6H2+H | $4.0 \times 10^{13}$ | 0.0 | 0.0 | Colket 86 | (7) |
| C6H2+H => C4H2+C2H | $9.3 \times 10^{14}$ | 0.0 | 15.1 | Colket 86 | (-7) |
| **Reactions of nC4H3 (CH≡CCH=CH•)** | | | | | |
| nC4H3 = C4H2+H (high pressure limit) | $1.0 \times 10^{14}$ | 0.0 | 36.0 | Miller 92[4] | (8) |
| (low pressure limit) | $1.0 \times 10^{14}$ | 0.0 | 30.0 | | |
| (Troe's coefficients) | /1.0   1.0 | | $1.0 \times 10^8$/ | | |
| C3H3+CH = nC4H3+H | $7.0 \times 10^{13}$ | 0.0 | 0.0 | Miller 92 | (9) |
| **nC4H3+H = iC4H3+H** | **$2.4 \times 10^{11}$** | **0.79** | **2.41** | **Wang97[6]** | **(10)** |
| **nC4H3+H = C4H2+H2** | **$6.0 \times 10^{12}$** | **0.0** | **0.0** | **Estimated[a]** | **(11)** |
| 2C2H2 = nC4H3+H | $1.0 \times 10^{12}$ | 0.0 | 64.1 | Leung 95 | (12) |
| C2H+C2H3 = nC4H3+H | $1.8 \times 10^{13}$ | 0.0 | 0.0 | Tsang 86 | (13) |
| **nC4H3+OH = C4H2+H2O** | **$1.5 \times 10^{13}$** | **0.0** | **0.0** | **Estimated[a]** | **(14)** |
| nC4H3+C2H2 = C6H4+H | $1.64 \times 10^9$ | 0.73 | 12.2 | Westmoreland 89[3] | (15) |
| nC4H3+C2H2 = lC6H4+H | 29.6 | 3.33 | 9.6 | Westmoreland 89 | (16) |
| nC4H3+C2H2 = lC6H5 | $1.73 \times 10^{11}$ | -0.41 | 4.0 | Westmoreland 89 | (17) |
| nC4H3+C2H2 = C6H5 | $3.33 \times 10^{24}$ | -3.89 | 9.2 | Westmoreland 89 | (18) |
| **Reactions of iC4H3 (CH≡C•C=CH2 ↔ •CH=C=C=CH2, resonnance stabilized radicals)** | | | | | |
| iC4H3 = nC4H3 | $1.5 \times 10^{13}$ | 0.0 | 67.8 | Leung 95 | (19) |
| iC4H3 = C4H2+H (high pressure limit) | $1.0 \times 10^{14}$ | 0.0 | 55.0 | Miller 92 | (20) |
| (low pressure limit) | $2.0 \times 10^{15}$ | 0.0 | 48.0/ | | |
| (Troe's coefficients) | /1.0   1.0 | | $1.0 \times 10^8$/ | | |
| C3H2+$^3$CH2 = iC4H3+H | $1.2 \times 10^{14}$ | 0.0 | 0.8 | Fournet99[8] | (21) |
| **iC4H3+H = 2C2H2** | **$2.4 \times 10^{19}$** | **-1.6** | **2.8** | **Wang97** | **(22)** |
| C3H3+CH = iC4H3+H | $7.0 \times 10^{13}$ | 0.0 | 0.0 | Miller 92 | (23) |
| **iC4H3+H = C4H2+H2** | **$1.2 \times 10^{13}$** | **0.0** | **0.0** | **Estimated[b]** | **(24)** |
| iC4H3+O = CH2CO+C2H | $2.0 \times 10^{13}$ | 0.0 | 0.0 | Miller 92 | (25) |
| **iC4H3+OH = C4H2+H2O** | **$3.0 \times 10^{13}$** | **0.0** | **0.0** | **Estimated[b]** | **(26)** |
| iC4H3+O2 = CH2CO+CHCO | $1.0 \times 10^{12}$ | 0.0 | 0.0 | Miller 92 | (27) |
| **iC4H3+C2H3 = 2C3H3** | **$4.0 \times 10^{12}$** | **0.0** | **0.0** | **Miller 92** | **(28)** |
| **iC4H3+C2H3 = lC6H5+H** | **$6.0 \times 10^{12}$** | **0.0** | **0.0** | **Miller 92** | **(29)** |
| **Reactions of C4H4 (CH≡CCH=CH2, vinylacetylene)** | | | | | |
| C3H3+$^3$CH2 = C4H4+H | $4.0 \times 10^{13}$ | 0.0 | 0.0 | Miller 92 | (30) |
| C2H3+C2H2 => C4H4+H | $2.0 \times 10^{13}$ | 0.0 | 25.1 | Douté 95[33] | (31) |
| C4H4+H = C2H3+C2H2 | $2.0 \times 10^{13}$ | 0.0 | 12.4 | Douté 95 | (32) |
| C2H4+C2H => C4H4+H | $1.2 \times 10^{13}$ | 0.0 | 0.0 | Tsang 86 | (33) |

| Reaction | A | n | Ea | Reference | # |
|---|---|---|---|---|---|
| **C4H4+H = nC4H3+H2** | **$2.0 \times 10^7$** | **2.0** | **15.5** | **Miller 92** | **(34)** |
| **C4H4+H = iC4H3+H2** | **$3.0 \times 10^7$** | **2.0** | **5.0** | **Miller 92** | **(35)** |
| C2H+C4H4 => C2H2+iC4H3 | $4.0 \times 10^{13}$ | 0.0 | 0.0 | Colket 86[34] | (36) |
| C2H2+iC4H3 => C2H+C4H4 | $3.0 \times 10^{13}$ | 0.0 | 27.9 | Colket 86 | (37) |
| C2H3+C4H4 => C2H4+nC4H3 | $5.0 \times 10^{11}$ | 0.0 | 16.3 | Colket 86 | (38) |
| nC4H3+C2H4 => C2H3+C4H4 | $3.5 \times 10^{11}$ | 0.0 | 13.4 | Colket 86 | (39) |
| C2H3+C4H4 => C2H4+iC4H3 | $5.0 \times 10^{11}$ | 0.0 | 16.3 | Colket 86 | (40) |
| iC4H3+C2H4 => C2H3+C4H4 | $1.3 \times 10^{11}$ | 0.0 | 24.1 | Colket 86 | (41) |
| **C4H4+C2H2=C6H5+H** | **$1.0 \times 10^9$** | **0.0** | **3.02** | **Benson92[35]** | **(42)** |
| C4H4+C2H3 = C6H6#+H | $1.9 \times 10^{12}$ | 0.0 | 2.5E3 | Lindstedt 96[17] | (43) |
| C4H4+O = aC3H4+CO | $3.0 \times 10^{13}$ | 0.0 | 1.8 | Leung 95 | (44) |
| C4H4+OH = nC4H3+H2O | $7.5 \times 10^6$ | 2.0 | 5.0 | Miller 92 | (45) |
| C4H4+OH = iC4H3+H2O | $1.0 \times 10^{71}$ | 2.0 | 2.0 | Miller 92 | (46) |
| aC3H4+aC3H4 = C2H4+C4H4 | $1.0 \times 10^{15}$ | 0.0 | 48.0 | Hidaka 89[36] | (47) |

**Reactions of nC4H5 (CH2=CHCH=CH•)**

| Reaction | A | n | Ea | Reference | # |
|---|---|---|---|---|---|
| nC4H5= H+C4H4 (high pressure limit) | $1.0 \times 10^{14}$ | 0.0 | 37.0 | Miller 92 | (48) |
| (low pressure limit) | $1.0 \times 10^{14}$ | 0.0 | 30.0 | | |
| nC4H5+H = C4H4+H2 | $1.5 \times 10^{13}$ | 0.0 | 0.0 | Wang 97 | (49) |
| nC4H5+H = iC4H5+H | $1.0 \times 10^{14}$ | 0.0 | 0.0 | Miller 92 | (50) |
| **nC4H5+CH3 =C5H8** | **$1.0 \times 10^{13}$** | **0.0** | **0.0** | **Estimated[c]** | **(51)** |
| nC4H5 = C2H2+C2H3 | $1.0 \times 10^{14}$ | 0.0 | 43.9 | Hidaka 96[37] | (52) |
| nC4H5+C2H2 = lC6H6+H | $1.17 \times 10^{-15}$ | 7.84 | 2.0 | Westmoreland 89 | (53) |
| nC4H5+C2H2 = C6H6#+H | $1.90 \times 10^7$ | 1.47 | 4.2 | Westmoreland 89 | (54) |
| nC4H5+C2H2 = lC6H7 | $8.74 \times 10^{12}$ | -1.27 | 3.6 | Westmoreland 89 | (55) |
| nC4H5+C2H2 = C6H7# | $1.96 \times 10^{19}$ | -3.35 | 5.2 | Westmoreland 89 | (56) |
| nC4H5+C2H3 = lC6H7+H | $8.28 \times 10^{-28}$ | 11.89 | 5.0 | Westmoreland 89 | (57) |
| nC4H5+C2H3 = lC6H8 | $2.90 \times 10^{15}$ | -0.78 | 1.0 | Westmoreland 89 | (58) |
| nC4H5+C2H3 = C6H8# | $5.50 \times 10^{15}$ | -1.67 | 1.5 | Westmoreland 89 | (59) |
| nC4H5+C2H3 = C6H6#+H2 | $2.80 \times 10^{-7}$ | 5.63 | -1.9 | Westmoreland 89 | (60) |
| nC4H5+OH = C4H4+H2O | $2.5 \times 10^{12}$ | 0.0 | 0.0 | Wang 97 | (61) |
| **nC4H5+O2 = C2H3CHO+CHO** | **$4.5 \times 10^{16}$** | **-1.39** | **1.0** | **Estimated[b]** | **(62)** |
| **nC4H5+O2 = C4H4+HO2** | **$4.5 \times 10^6$** | **1.61** | **-0.4** | **Estimated[b]** | **(63)** |

**Reactions of iC4H5 (CH2=CH•C=CH2↔• CH2CH=C=CH2, resonnance stabilized radicals)**

| Reaction | A | n | Ea | Reference | # |
|---|---|---|---|---|---|
| iC4H5 = nC4H5 | $1.5 \times 10^{13}$ | 0.0 | 67.8 | Leung 95 | (64) |
| iC4H5= H+C4H4 (high pressure limit) | $1.0 \times 10^{14}$ | 0.0 | 50.0 | Miller 92 | (65) |
| (low pressure limit) | $2.0 \times 10^{15}$ | 0.0 | 42.0 | | |
| iC4H5+H = C4H4+H2 | $3.0 \times 10^{13}$ | 0.0 | 0.0 | Wang 97 | (66) |
| 2C2H3 = iC4H5+H | $1.5 \times 10^{30}$ | -4.95 | 13.7 | Wang 97 | (67) |
| **iC4H5+CH3 =iC5H8** | **$1.0 \times 10^{13}$** | **0.0** | **0.0** | **Estimated[c]** | **(68)** |
| iC4H5+OH = C4H4+H2O | $5.5 \times 10^{12}$ | 0.0 | 0.0 | Wang 97 | (69) |
| **iC4H5+O2 = CH2CHCO+HCHO** | **$4.5 \times 10^{16}$** | **-1.39** | **1.0** | **Estimated[b]** | **(70)** |
| **iC4H5+O2 = C4H4+HO2** | **$4.5 \times 10^6$** | **1.61** | **-0.4** | **Estimated[b]** | **(71)** |

**Reactions of 1,3-C4H6 (CH2=CHCH=CH2, 1,3-butadiene)**

| Reaction | A | n | Ea | Reference | # |
|---|---|---|---|---|---|
| C2H3+C2H3 = 1,3-C4H6 | $9.8 \times 10^{14}$ | -0.5 | 0.0 | Hidaka 96 | (72) |
| 1,3-C4H6 = C4H4+H2 | $2.5 \times 10^{15}$ | 0.0 | 94.7 | Hidaka 96 | (73) |
| 1,3-C4H6 = iC4H5+H | $1.4 \times 10^{15}$ | 0.0 | 98.0 | Hidaka 96 | (74) |
| C2H4+C2H3 = 1,3-C4H6+H | $5.0 \times 10^{11}$ | 0.0 | 7.3 | Tsang 86 | (75) |
| 1,3-C4H6+H = nC4H5+H2 | $1.3 \times 10^6$ | 2.53 | 12.2 | Wang 97 | (76) |
| 1,3-C4H6+H = iC4H5+H2 | $6.6 \times 10^5$ | 2.53 | 9.2 | Wang 97 | (77) |
| **1,3-C4H6+H = 1C4H7-1** | **$2.6 \times 10^{13}$** | **0.0** | **3.2** | **Estimated[d]** | **(78)** |
| **1,3-C4H6+H = 1C4H7Y** | **$2.6 \times 10^{13}$** | **0.0** | **1.6** | **Estimated[d]** | **(79)** |
| 1,3-C4H6+CH3 = nC4H5+CH4 | $7.0 \times 10^{13}$ | 0.0 | 18.5 | Wu 87 | (80) |
| 1,3-C4H6+CH3 = iC4H5+CH4 | $7.0 \times 10^{13}$ | 0.0 | 15.5 | Fournet99 | (81) |
| **1,3-C4H6+CH3 = C5H9Y** | **$6.3 \times 10^{10}$** | **0.0** | **7.5** | **Perrin88[38]** | **(82)** |

| Reaction | A | n | E | Source | # |
|---|---|---|---|---|---|
| **1,3-C4H6+CH3 = iC5H9** | **1.8x10¹¹** | **0.0** | **8.0** | **Estimated[d]** | **(83)** |
| 1,3-C4H6+C2H3 = nC4H5+C2H4  5 | 5.0x10¹⁴ | 0.0 | 22.8 | Hidaka 96 | (84) |
| 1,3-C4H6+C2H3 = iC4H5+C2H4 | 5.0x10¹⁴ | 0.0 | 19.8 | Fournet99 | (85) |
| 1,3-C4H6+C2H2 = C6H8# | 2.3x10¹² | 0.0 | 35.0 | Westmoreland 89 | (86) |
| 1,3-C4H6+C2H3 = C6H8#+H | 2.28x10¹² | -0.24 | 9.9 | Westmoreland 89 | (87) |
| 1,3-C4H6+C2H3 = lC6H8+H | 1.0x10¹⁰ | 1.05 | 14.0 | Westmoreland 89 | (88) |
| 1,3-C4H6+C2H3 = lC6H9 | 5.48x10²⁸ | -5.31 | 9.3 | Westmoreland 89 | (89) |
| 1,3-C4H6+C2H3 = C6H9# | 1.64x10²⁹ | -6.12 | 9.6 | Westmoreland 89 | (90) |
| 1,3-C4H6+C2H4 = C6H10# | 2.3x10¹⁰ | 0.0 | 27.0 | Westmoreland 89 | (91) |
| 1,3-C4H6+O = aC3H5+H+CO | 6.0x10⁸ | 1.45 | 0.9 | Fournet99 | (92) |
| 1,3-C4H6+OH = nC4H5+H2O | 6.2x10⁶ | 2.0 | 3.4 | Wang 97 | (91) |
| 1,3-C4H6+OH = iC4H5+H2O | 3.1x10⁶ | 2.0 | 0.4 | Wang 97 | (94) |
| 1,3-C4H6+OH = aC3H5+HCHO | 2.8x10¹² | 0.0 | -0.9 | Lindstedt 96 | (95) |
| 1,3-C4H6+OH = CH3CHO+C2H3 | 5.6x10¹² | 0.0 | -0.9 | Lindstedt 96 | (96) |
| 1,3-C4H6+O2 = iC4H5+HO2 | 4.0x10¹³ | 0.0 | 57.9 | Leung 95 | (97) |
| 1,3-C4H6+C3H3 = nC4H5+aC3H4 | 1.0x10¹³ | 0.0 | 22.5 | Hidaka 96 | (98) |
| 1,3-C4H6+C3H3 = iC4H5+aC3H4 | 1.0x10¹³ | 0.0 | 19.5 | Fournet99 | (99) |

**Reactions of 1,2-C4H6 (CH2=C=CHCH3, 1,2-butadiene)**

| Reaction | A | n | E | Source | # |
|---|---|---|---|---|---|
| 1,2-C4H6 = 1,3-C4H6 | 3.0x10¹³ | 0.0 | 65.0 | Hidaka 96 | (100) |
| **1,2-C4H6 = C3H3+CH3** | **7.3x10¹⁴** | **0.0** | **75.4** | **Estimated[c]** | **(101)** |
| 1,2-C4H6 = iC4H5+H | 4.2x10¹⁵ | 0.0 | 92.6 | Leung 95 | (102) |
| 1,2-C4H6+H = C2H3+C2H4 | 4.0x10¹¹ | 0.0 | 0.0 | Leung 95 | (103) |
| **1,2-C4H6+H = 2C4H7** | **1.3x10¹³** | **0.0** | **1.6** | **Estimated[d]** | **(104)** |
| **1,2-C4H6+H = 1C4H7Y** | **1.2x10¹¹** | **0.69** | **3.0** | **Estimated[e]** | **(105)** |
| **1,2-C4H6+H = 1C4H7T** | **1.3x10¹³** | **0.0** | **3.2** | **Estimated[d]** | **(106)** |
| 1,2-C4H6+H = iC4H5+H2 | 1.7x10⁵ | 2.5 | 2.5 | Fournet99 | (107) |
| 1,2-C4H6+CH3 = iC4H5+CH4 | 2.2 | 3.5 | 5.7 | Fournet99 | (108) |
| 1,2-C4H6+C2H5 = iC4H5+C2H6 | 2.2 | 3.5 | 6.6 | Fournet99 | (109) |
| 1,2-C4H6+O = iC4H5+OH | 1.7x10¹¹ | 0.7 | 5.9 | Fournet99 | (110) |
| 1,2-C4H6+OH = iC4H5+H2O | 3.1x10⁶ | 2.0 | -0.3 | Fournet99 | (111) |
| 1,2-C4H6+HO2 = iC4H5+H2O2 | 9.6x10³ | 2.6 | 13.9 | Fournet99 | (112) |

**Reactions of 1C4H7-1 (CH2=CH-CH2-CH2•)**

| Reaction | A | n | E | Source | # |
|---|---|---|---|---|---|
| 1C4H7-1 = 1C4H7Y | 3.3x10⁹ | 1.0 | 39.1 | Estimated[f] | (113) |
| 1C4H7-1 = 1C4H7V | 3.3x10⁹ | 1.0 | 20.7 | Estimated[f] | (114) |
| 1C4H7-1 = C2H4+C2H3 | 2.0x10¹³ | 0.0 | 35.5 | Estimated[g] | (115) |
| 1C4H7-1+H=C4H8 | 2.0x10¹³ | 0.0 | 0.0 | Estimated[d] | (116) |
| 1C4H7-1+CH3=C5H10 | 1.0x10¹⁴ | 0.0 | 0.0 | Estimated[h] | (117) |

**Reactions of 1C4H7Y (CH2=CH-CH•-CH3↔ •CH2-CH=CH-CH3, resonnance stabilized radicals)**

| Reaction | A | n | E | Source | # |
|---|---|---|---|---|---|
| 1C4H7Y+H = C4H8Y | 2.0x10¹³ | 0.0 | 0.0 | Estimated[d] | (118) |
| 1C4H7Y+H = 1,3-C4H6+H2 | 0.9x10¹³ | 0.0 | 0.0 | Estimated[d] | (119) |
| 1C4H7Y+H = 1,2-C4H6+H2 | 0.9x10¹³ | 0.0 | 0.0 | Estimated[d] | (120) |
| 1C4H7Y+HO2=OH+C2H3CHO+CH3 | 1.0x10¹⁵ | -0.8 | 0.0 | Estimated[g] | (121) |
| 1C4H7Y+CH3 = iC5H10 | 0.5x10¹³ | 0.0 | 0.0 | Estimated[d] | (122) |

**Reactions of 1C4H7V (CH3-CH2-CH=CH•)**

| Reaction | A | n | E | Source | # |
|---|---|---|---|---|---|
| 1C4H7V = 1C4H7Y | 1.9x10¹⁰ | 1.0 | 36.3 | Estimated[f] | (123) |
| 1C4H7V = C2H5+C2H2 | 2.0x10¹³ | 0.0 | 33.0 | Estimated[g] | (124) |

**Reactions of 1C4H7T (CH2=C•-CH2-CH3)**

| Reaction | A | n | E | Source | # |
|---|---|---|---|---|---|
| 1C4H7T = 1C4H7-1 | 3.3x10⁹ | 1.0 | 43.3 | Estimated[f] | (125) |
| 1C4H7T = 1C4H7Y | 2. 0x10¹³ | 0.0 | 47.0 | Estimated[g] | (126) |
| 1C4H7T =CH3+aC3H4 | 2.0x10¹³ | 0.0 | 32.5 | Estimated[g] | (127) |

**Reactions of 2C4H7 (CH3-C•=CH-CH3)**

| | | | | | |
|---|---|---|---|---|---|
| 2C4H7 = 1C4H7Y | $2.9 \times 10^{10}$ | 1.0 | 37.8 | Estimated$^f$ | (128) |
| 2C4H7 = CH3+pC3H4 | $2.0 \times 10^{13}$ | 0.0 | 31.5 | Estimated$^g$ | (129) |

**Reactions of C5H7Y (CH2=CH-CH=CH-CH2• ↔ CH2=CH-CH•-CH=CH2, resonnance stabilized radicals)**

| | | | | | |
|---|---|---|---|---|---|
| C5H7Y+H=C5H8 | $1.0 \times 10^{14}$ | 0.0 | 0.0 | Estimated$^h$ | (130) |

**Reactions of iC5H7Y (CH2=C (CH2•)-CH=CH2, resonnance stabilized radicals)**

| | | | | | |
|---|---|---|---|---|---|
| iC5H7Y=C2H3+aC3H4 | $2.0 \times 10^{13}$ | 0.0 | 50.0 | Estimated$^e$ | (131) |
| iC5H7Y+H=iC5H8 | $1.0 \times 10^{14}$ | 0.0 | 0.0 | Estimated$^h$ | (132) |

**Reactions of C5H8 (CH2=CH-CH=CH-CH3)**

| | | | | | |
|---|---|---|---|---|---|
| C5H8+H=C5H7Y+H2 | $1.7 \times 10^{5}$ | 2.50 | 2.5 | Estimated$^g$ | (133) |
| C5H8+CH3=C5H7Y+CH4 | 2.2 | 3.50 | 5.7 | Estimated$^g$ | (134) |
| C5H8+OH=H2O+C5H7Y | $3.0 \times 10^{6}$ | 2.0 | -0.3 | Estimated$^g$ | (135) |

**Reactions of iC5H8 (CH2=C (CH3)-CH=CH2)**

| | | | | | |
|---|---|---|---|---|---|
| iC5H8+H=iC5H7Y+H2 | $1.7 \times 10^{5}$ | 2.50 | 2.5 | Estimated$^g$ | (136) |
| iC5H8+CH3=iC5H7Y+CH4 | 2.2 | 3.50 | 5.7 | Estimated$^g$ | (137) |
| iC5H8+OH=H2O+iC5H7Y | $3.0 \times 10^{6}$ | 2.0 | -0.3 | Estimated$^g$ | (138) |

**Reactions of C5H9Y (CH2=CH-CH•-CH2-CH3 ↔ •CH2-CH=CH-CH2-CH3, resonnance stabilized radicals)**

| | | | | | |
|---|---|---|---|---|---|
| C5H9Y=H+C5H8 | $3.0 \times 10^{13}$ | 0.0 | 50.5 | Estimated$^e$ | (139) |
| C5H9Y+H= C5H10 | $2.0 \times 10^{13}$ | 0.0 | 0.0 | Estimated$^d$ | (140) |
| C5H9Y+H= C5H8+H2 | $1.8 \times 10^{13}$ | 0.0 | 0.0 | Estimated$^d$ | (141) |

**Reactions of iC5H9 (CH2•-CH (CH3)-CH=CH2)**

| | | | | | |
|---|---|---|---|---|---|
| iC5H9=iC5H9Y | $1.7 \times 10^{9}$ | 1.0 | 38.1 | Estimated$^f$ | (142) |
| iC5H9=C3H6+C2H3 | $2.0 \times 10^{13}$ | 0.0 | 35.5 | Estimated$^e$ | (143) |
| iC5H9=iC5H8+H | $1.6 \times 10^{13}$ | 0.0 | 34.3 | Estimated$^e$ | (144) |

**Reactions of iC5H9Y (CH2=C•(CH3)-CH=CH2 ↔ CH2=C(CH3)=CH-CH2•, resonnance stabilized radicals)**

| | | | | | |
|---|---|---|---|---|---|
| iC5H9Y= iC5H8+H | $1.0 \times 10^{13}$ | 0.0 | 51.5 | Estimated$^g$ | (145) |
| iC5H9Y+H= iC5H10 | $2.0 \times 10^{13}$ | 0.0 | 0.0 | Estimated$^d$ | (146) |
| iC5H9Y+H= iC5H8+H2 | $1.8 \times 10^{13}$ | 0.0 | 0.0 | Estimated$^d$ | (147) |
| iC5H9Y+aC3H5 = C3H6+iC5H8 | $8.4 \times 10^{10}$ | 0.0 | 0.0 | Estimated$^d$ | (148) |
| iC5H9Y+ iC5H9Y = iC5H10+iC5H8 | $8.4 \times 10^{10}$ | 0.0 | 0.0 | Estimated$^d$ | (149) |

**Reactions of C5H10 (CH2=CH-CH2-CH2-CH3)**

| | | | | | |
|---|---|---|---|---|---|
| C5H10+H=C5H9Y+H2 | $5.4 \times 10^{4}$ | 2.5 | -1.9 | Estimated$^g$ | (150) |
| C5H10+CH3=C5H9Y+CH4 | $1.0 \times 10^{11}$ | 0.0 | 7.3 | Estimated$^g$ | (151) |
| C5H10+OH=C5H9Y+H2O | $3.0 \times 10^{6}$ | 2.0 | -1.5 | Estimated$^g$ | (152) |

**Reactions of iC5H10 (CH3-CH (CH3)-CH=CH2)**

| | | | | | |
|---|---|---|---|---|---|
| iC5H10+H=iC5H9Y+H2 | $2.5 \times 10^{4}$ | 2.5 | -1.9 | Estimated$^e$ | (153) |
| iC5H10+CH3=iC5H9Y+CH4 | $5.0 \times 10^{10}$ | 0.0 | 5.6 | Estimated$^e$ | (154) |
| iC5H10+OH=iC5H9Y+H2O | $1.3 \times 10^{6}$ | 2.0 | -2.6 | Estimated$^e$ | (155) |

**Reactions added for 1,4-cyclohexadiene and benzyne**

| Reaction | A | n | Ea | Reference | # |
|---|---|---|---|---|---|
| C6H8#=C6H6#+H2 | 1.28E28 | -4.94 | 49.3 | Ellis66[44] | (156) |
| C6H8#+H=H2+C6H7# | 1.1E05 | 2.5 | -1.9 | Dayma03[45] | (157) |
| C6H8#+OH=H2O+C6H7# | 6.0E6 | 2.0 | -1.52 | Dayma03 | (158) |
| C6H4#+H=C6H5# | 3.0E17 | 0.0 | 36.3 | Wang97 | (159) |

___________________________________________________________________

[a] : Rate constant considered as half of that of the similar reaction on the case of vinyl radicals [20].

[b] : Rate constant taken equal as that of the similar reaction on the case of vinyl radicals [20].

[c] : Rate constant of this unimolecular initiation calculated by the modified collision theory at 1500 K using software KINGAS [39].

[d] : Rate constant estimated by analogy with the values proposed by Tsang [40] for propene or allyl radicals, taking into account the number of bonds for additions.

[e] : Rate constant of this reaction estimated according to the correlations proposed by Heyberger [41].

[f] : Rate constant of this isomerization estimated as explained in the text and in Table II.

[g] : Rate constant of this reaction estimated according to the correlations for linear alkenes proposed by Heyberger et al. [42].

[h] : Rate constant taken equal to that of the recombination of •H atoms with alkyl radicals as proposed by Allara et al. [43].

**TABLE II: ESTIMATED RATE COEFFICIENTS FOR ISOMERIZATIONS OF BUTENYL AND PENTENYL RADICALS**

A-factors are given in s$^{-1}$ and energy in kcal/mol.

| N° in Table 1 | Structure of the transition state | $\Delta n_{int.\ rot.}$ | Reaction path degeneracy | A/T | $E_{abst.}$ [41] | $E_{cycle}$ [25] | E |
|---|---|---|---|---|---|---|---|
| 113 | | -2 | 2 | **3.3x10$^9$** | 6.5 (secondary allylic H-atom) | 27.6+5$^a$ (Unsaturated C$_3$ cycle) | **39.1** |
| 114 | | -2 | 2 | **3.3x10$^9$** | 14.8 (secondary vinylic H-atom) | 5.9 (Unsaturated C$_5$ cycle) | **20.7** |
| 123 | | -1 | 2 | **1.9x10$^{10}$** | 6.5 (secondary allylic H-atom) | 29.8 (Unsaturated C$_4$ cycle) | **36.3** |
| 125 | | -2 | 3 | **3.3x10$^9$** | 13.5 (primary alkylic H-atom) | 27.2$^b$ | **40.7** |
| 128 | | -1 | 3 | **2.9x10$^{10}$** | 8.0 (primary allylic H-atom) | 29.8 (Unsaturated C$_4$ cycle) | **37.8** |
| 142 | | -2 | 1 | **1.7x10$^9$** | 5.0 (tertiary allylic H-atom) | 27.6+5$^a$ (Unsaturated C$_3$ cycle) | **38.1** |

$^a$ : A correction of 5 kcal/mol is applied to take into account the difference in the ring strain energy between a C-H-C and the reference C$_3$ ring.
$^b$ : The ring strain energy is that of methylcyclobutane [41].

**FIGURE CAPTIONS**

Figure 1: Typical chromatogram of $C_1$-$C_6$ compounds obtained at a distance of 0.38 cm from the burner (oven temperature program: 313 K during 22 min, then a rise of 1 K/min until 523 K).

Figure 2: Temperature profiles in the three flames: experimental measurements performed without and with the sampling probe and profile used for simulation.

Figure 3: Profiles of the mole fractions of oxygen and $C_1$ species. Points are experiments and lines simulations. Full lines correspond to the flame seeded with 1,3-butadiene and broken lines to a simulated flame of pure methane at $\Phi=1.8$ (see text), but cannot be seen in fig. 3a.

Figure 4: Profiles of the mole fractions of hydrogen and $C_2$ species. Points are experiments and lines simulations. Full lines correspond to the flame seeded with 1,3-butadiene and the broken lines to a simulated flame of pure methane at $\Phi=1.8$ (see text).

Figure 5: Profiles of the mole fractions $C_3$ species and aromatic compounds. Points are experiments and lines simulations. Full lines correspond to the flame seeded with 1,3-butadiene and the broken lines to a simulated flame of pure methane at $\Phi=1.8$ (see text).

Figure 6: Profiles of the mole fractions $C_4$ species. Points are experiments and lines simulations.

Figure 7: Profiles of the mole fractions $C_5$ species. Points are experiments and lines simulations.

Figure 8: Flow rate analysis for the consumption of the 1,3-butadiene for a distance of 0.27 cm from the burner corresponding to a temperature of 980 K and a conversion of 47 % the $C_4$ reactant.

Figure 9: Modified simulated profiles of the mole fractions vinylacetylene and benzene in the flame doped by propyne and allene. Points are experiments [7], thin lines previous simulations and thick lines new simulations.

Figure 10: Flow rate analysis for the formation and consumption of phenyl radicals for a distance of 0.42 cm from the burner corresponding to a temperature of 1480 K, a conversion of 87% of the $C_4$ reactant and close to the peak of benzene profile.

Figure 1

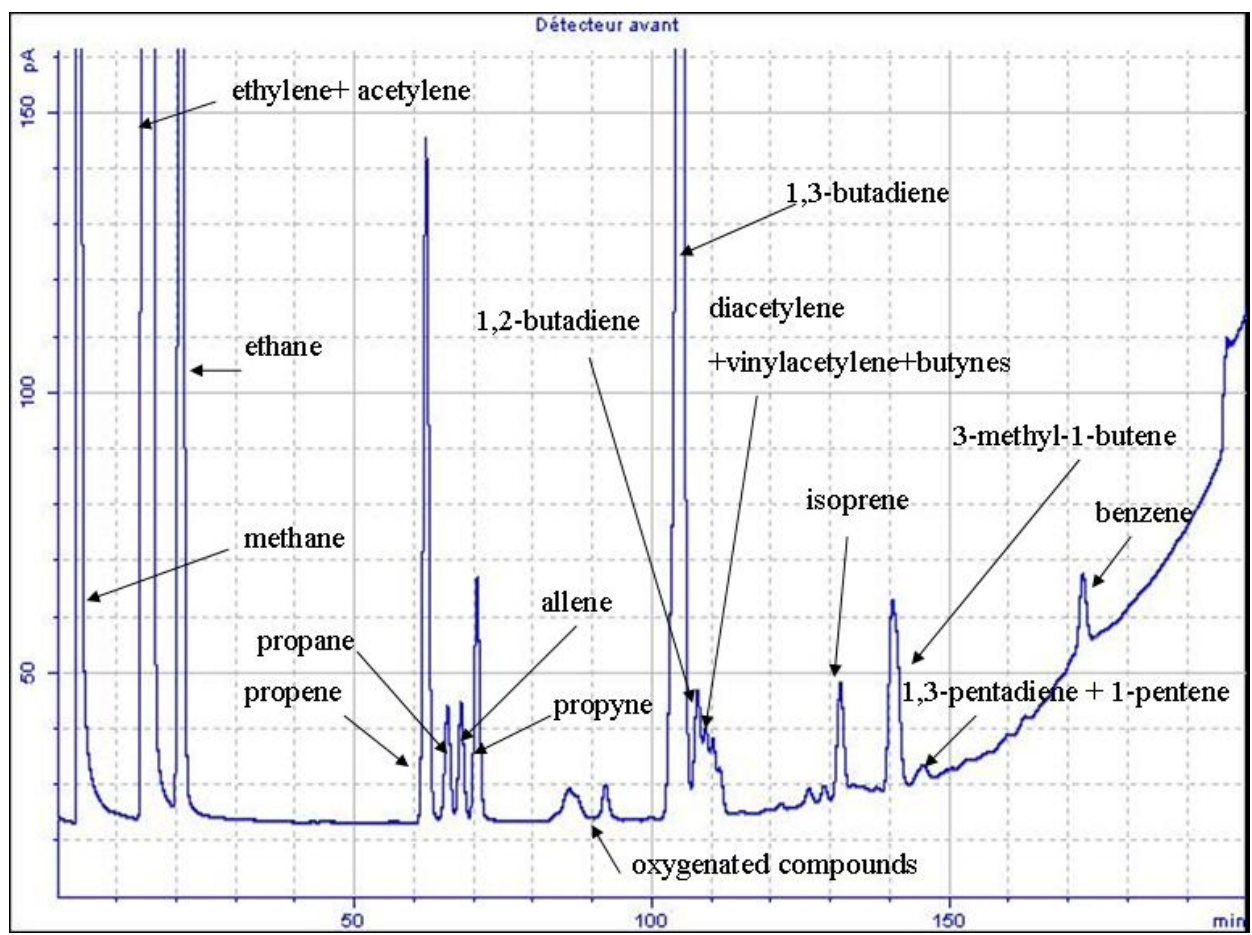

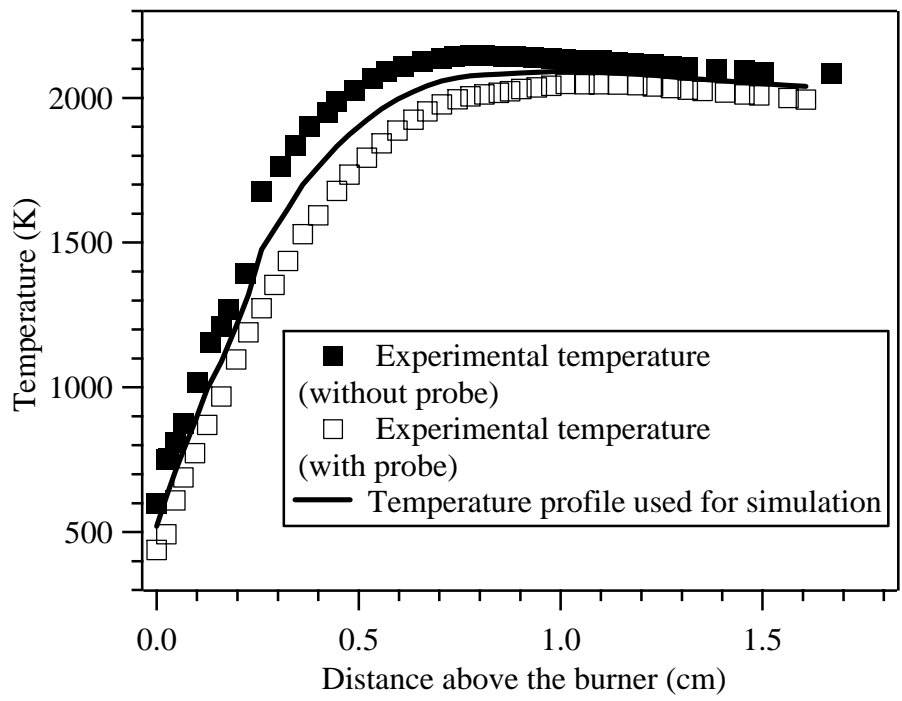

Figure 3

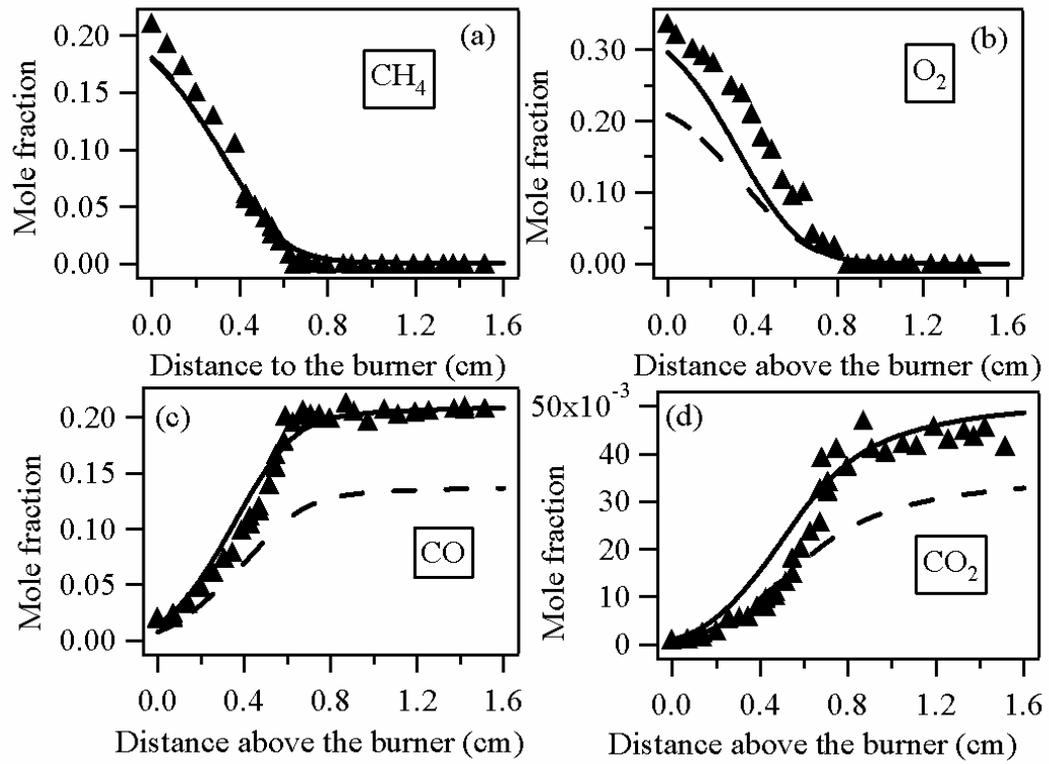

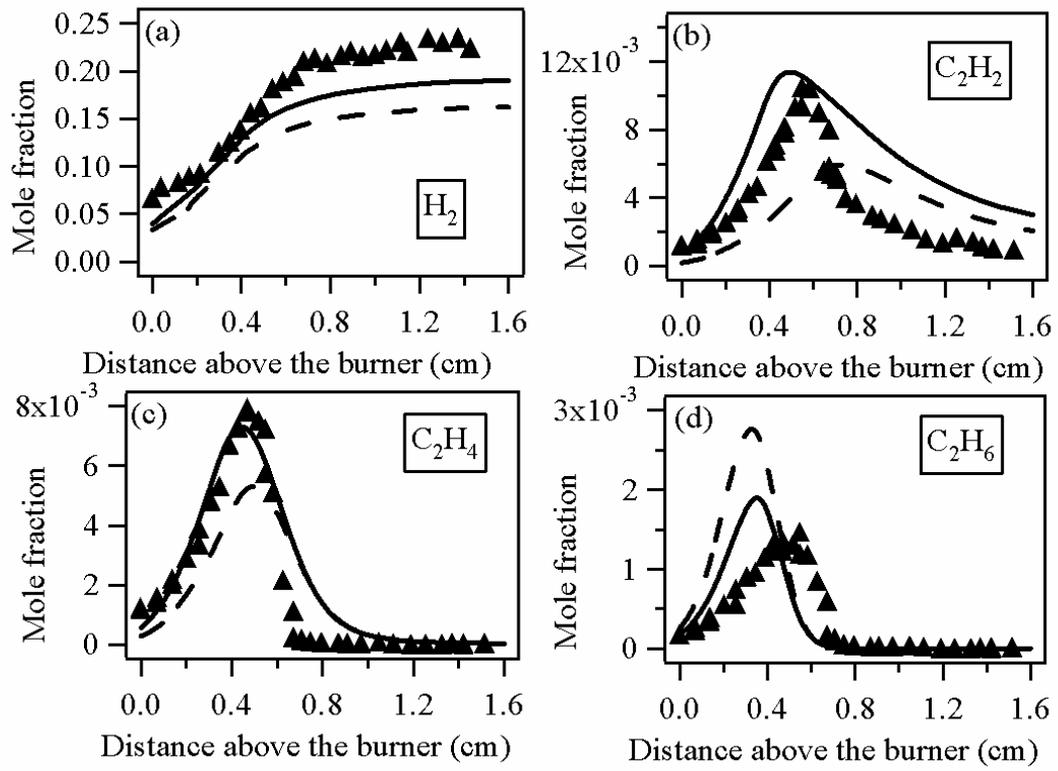

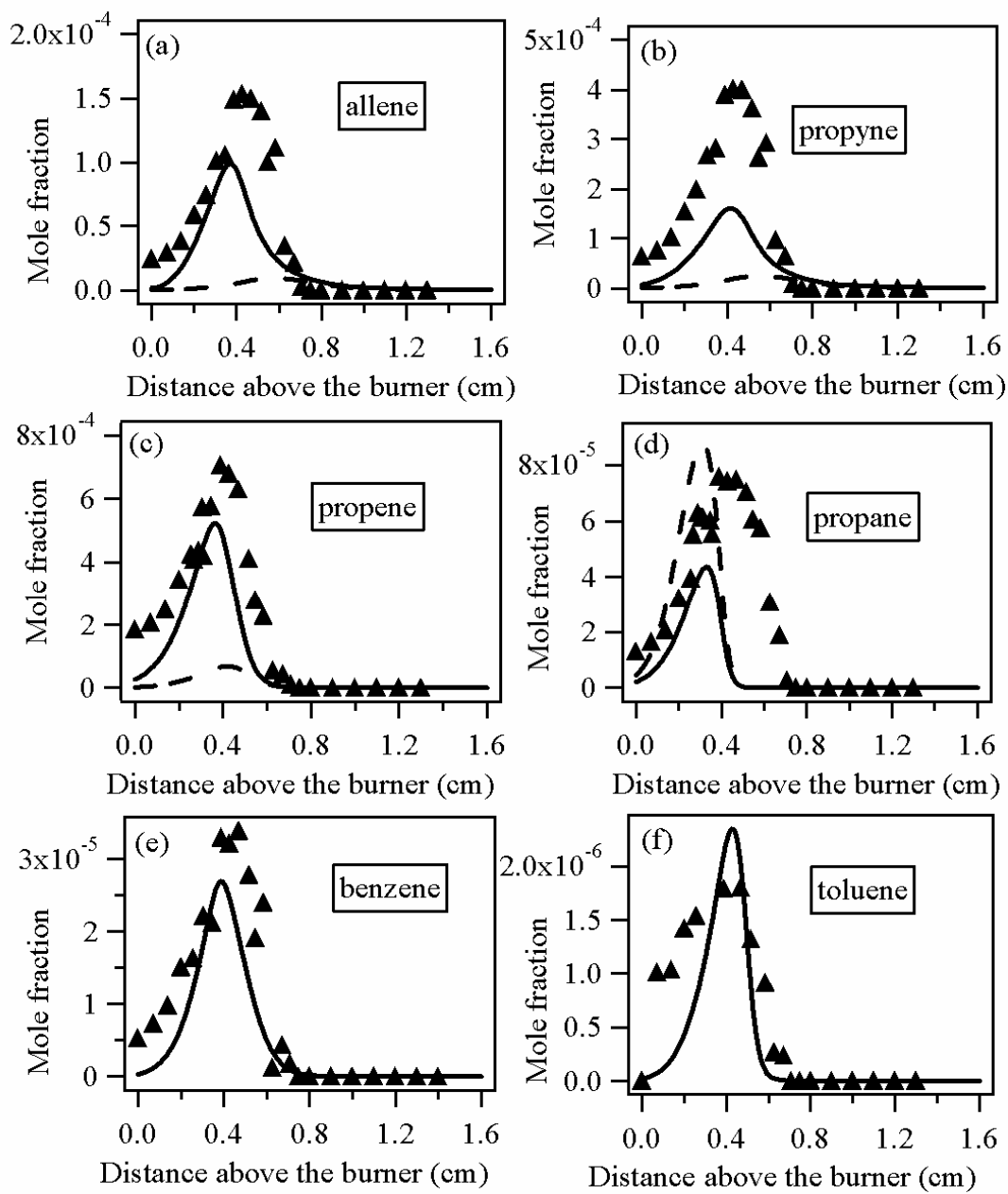

Figure 5

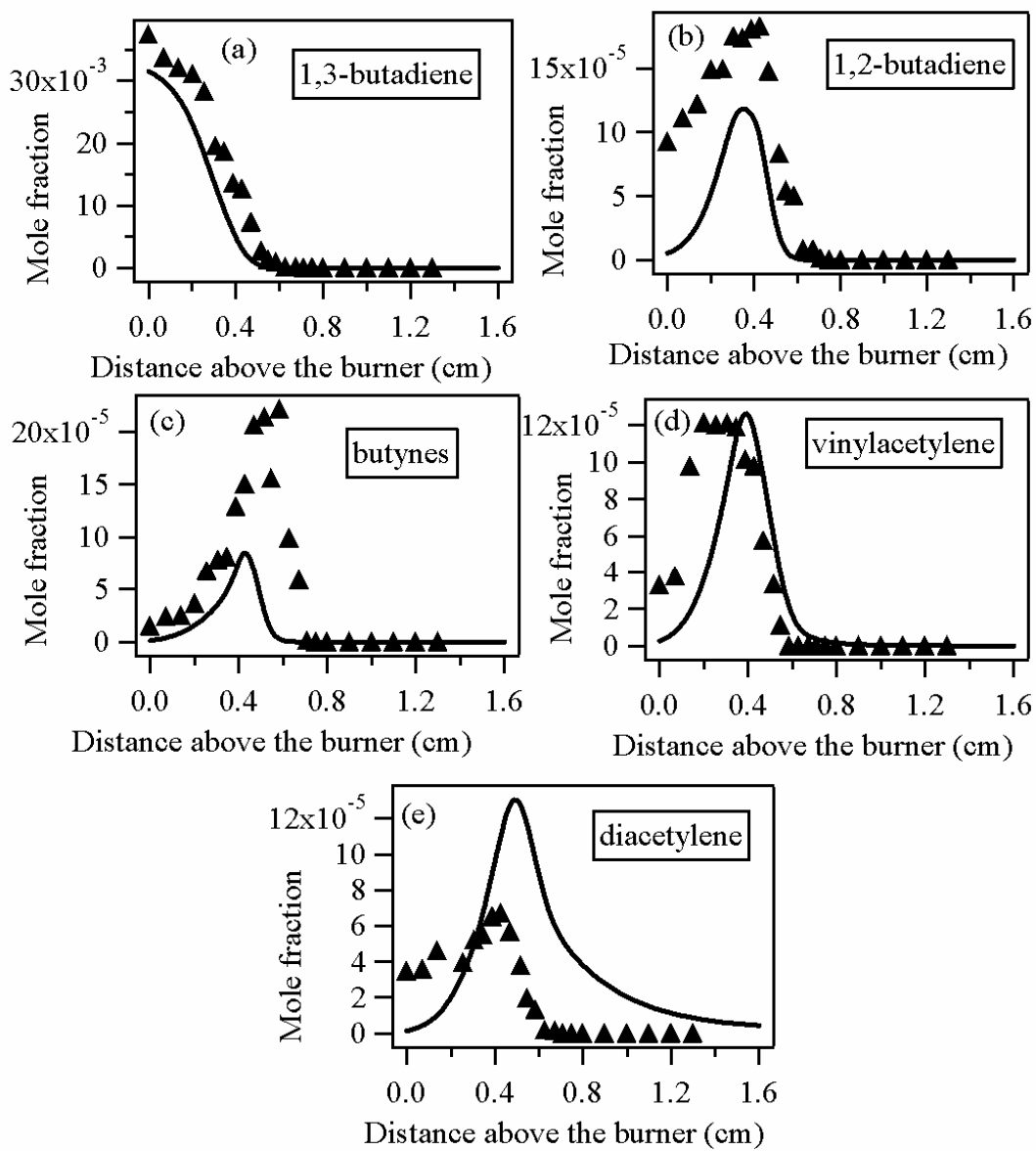

Figure 6

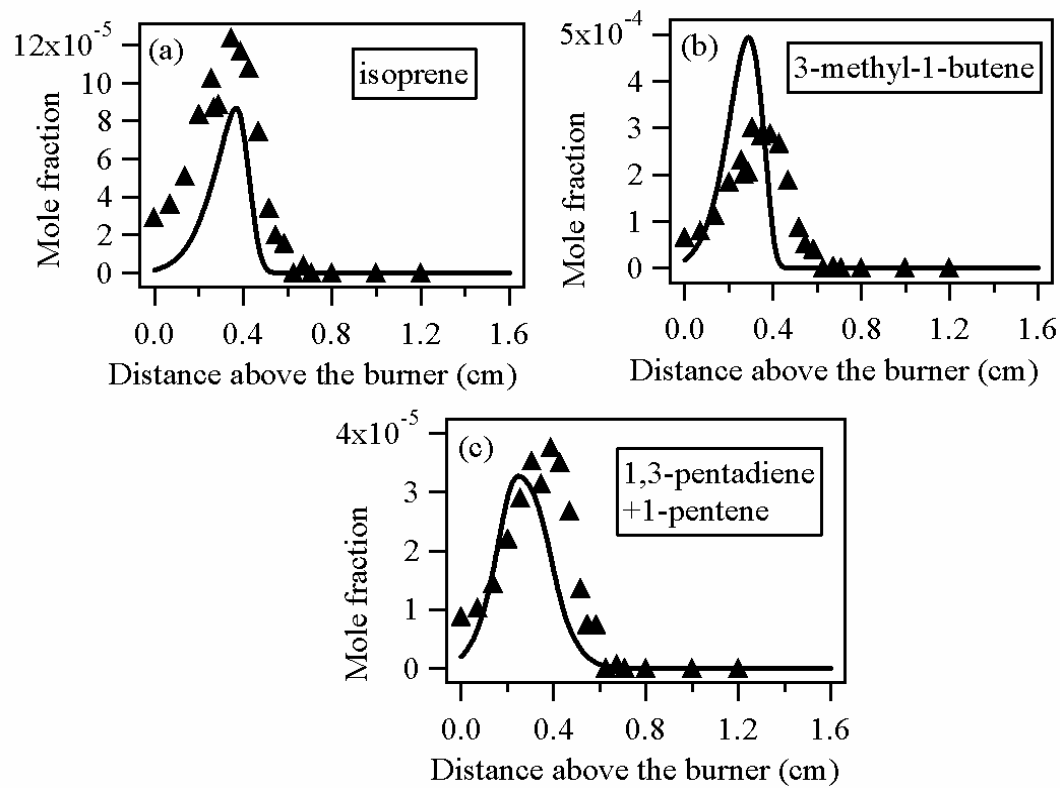

Figure 7

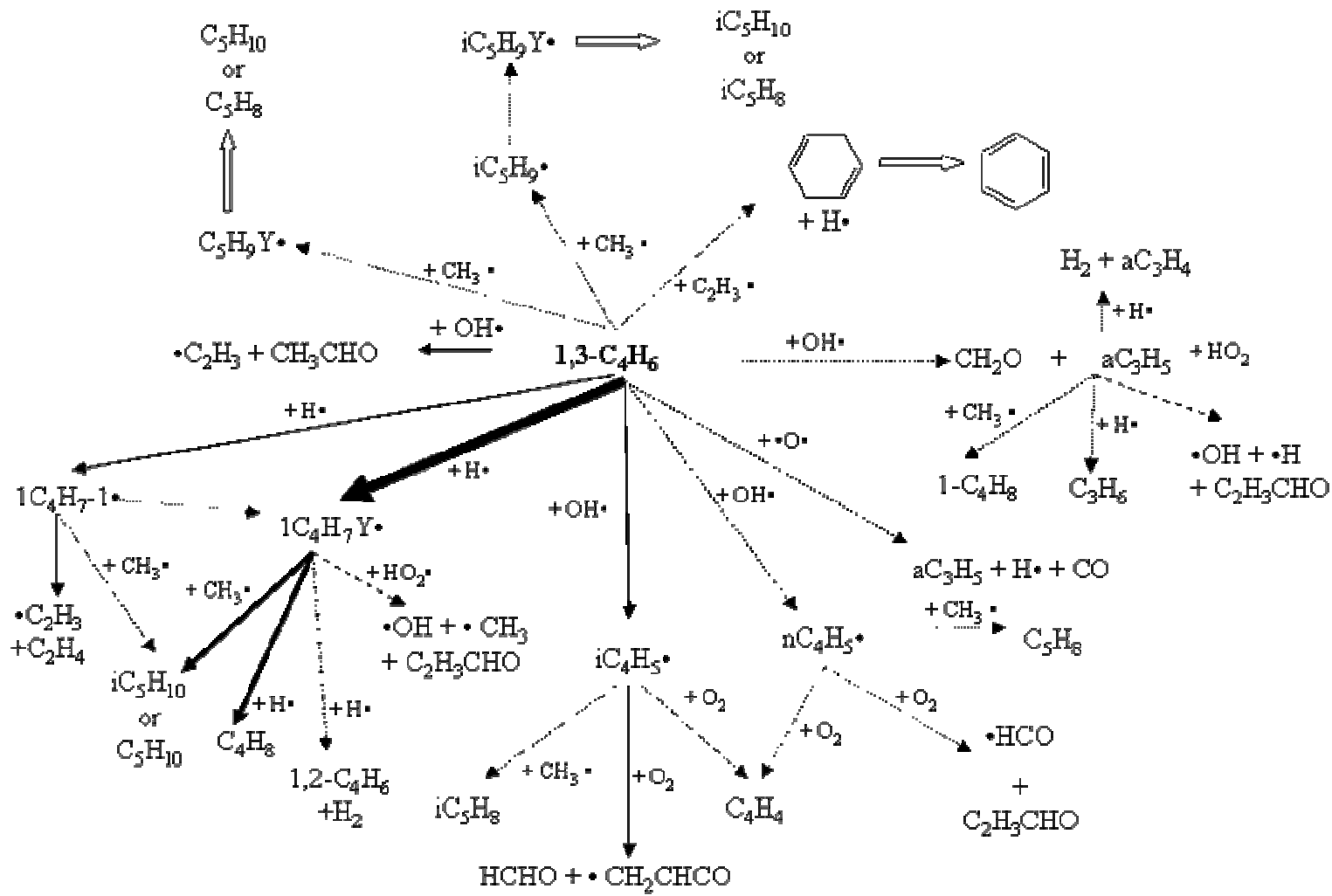

Figure 8

Figure 9

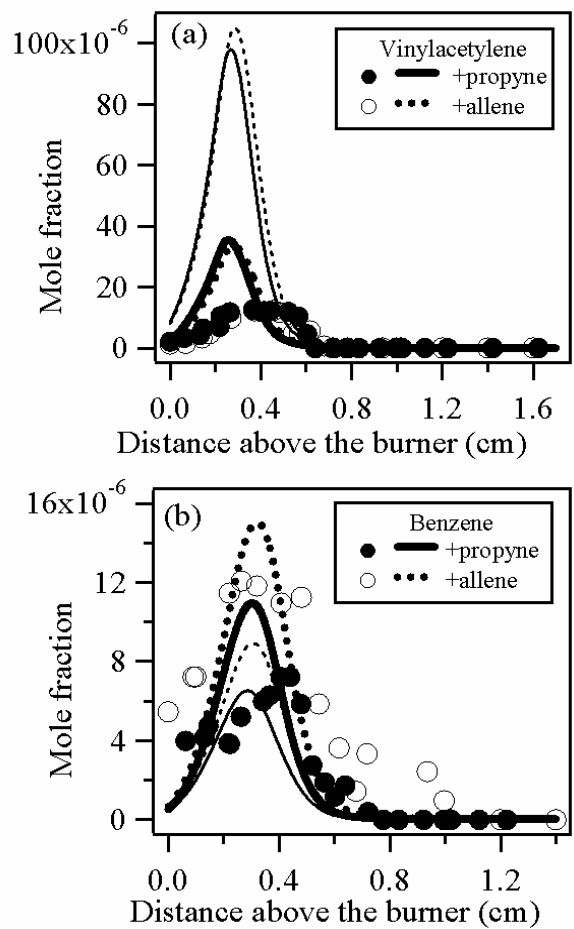